\documentclass[twocolumn,superscriptaddress,nofootinbib,floatfix,aps,prb,citeautoscript,longbibliography]{revtex4-2}
\usepackage[utf8]{inputenc}
\usepackage[T1]{fontenc}
\usepackage{lineno}

\usepackage{hyperref}
\usepackage{graphicx}%
\usepackage{booktabs}
\usepackage{color}
\usepackage{array}
\usepackage{dcolumn}
\usepackage{bm}
\usepackage{multirow}
\usepackage[version=4]{mhchem}
\usepackage{chemformula}
\let\ce\ch
\setchemformula{kroeger-vink=true}

\usepackage{cleveref}
\usepackage{float}
\usepackage{eso-pic}

\makeatletter
    \renewcommand\@make@capt@title[2]{%
     \@ifx@empty\float@link{\@firstofone}{\expandafter\href\expandafter{\float@link}}%
      {{#1}}\@caption@fignum@sep#2\quad}%
\makeatother
   
\makeatletter 
\renewcommand{\fnum@figure}{FIG.~\thefigure}
\makeatother

\newcommand{\mer}{$m_\text{e}$}
\newcommand{\me}{$m^*_\text{e}$}
\newcommand{\mh}{$m^*_\text{h}$}
\newcommand{\El}{$E_\text{VI}$}
\newcommand{\Eh}{$E_\text{IC}$}

\usepackage[breakable, theorems, skins]{tcolorbox}

\newcommand{\ehull}{$E_\text{hull}$}
\newcommand{\alex}{\textsc{alexandria}}

\begin{document}
\newcommand{\bochum}{Research Center Future Energy Materials and Systems of the University Alliance Ruhr and Interdisciplinary Centre for Advanced Materials Simulation, Ruhr University Bochum, Universitätsstraße 150, D-44801 Bochum, Germany}

\newcommand{\wvu}{Department of Physics, West Virginia University, Morgantown, WV 26506.}
\newcommand{\lux}{Department of Physics and Materials Science, University of Luxembourg, 162a avenue de la Fa\"iencerie, L-1511 Luxembourg, Luxembourg}

\title{Mapping the Phase Space of Cubic Chiral Semiconductors: High-Throughput Search, Stability, and Prototype Families}
\author{Pierre-Paul De Breuck}
\affiliation{\bochum}
\author{Hai-Chen Wang}
\affiliation{\bochum}
\author{Reynaldo Mares-Gil}
\affiliation{\wvu}
\author{Miguel A. L. Marques}
\affiliation{\bochum}
\author{Aldo H. Romero}
\affiliation{\wvu}
\affiliation{\lux}
\date{\today}

\begin{abstract}
Chiral crystals show unique electronic, magnetic, and optical characteristics that make them attractive for topological and spintronic applications. Nevertheless, the limited set of identified chiral semiconductors, especially those with cubic symmetry, has so far constrained their technological application. In this work, we perform a high-throughput computational search for semiconducting chiral compounds in the cubic space group $P4_332$ (No. 212). By integrating machine-learning techniques with density functional theory, we obtain 24\,227 cubic chiral structures containing up to five elements, and identify 368 thermodynamically stable semiconductors on the convex hull and 3\,048 within 50~meV/atom. We analyze their structural motifs, stability, and electronic band structures, highlighting recurring prototype families, several of which show intrinsic intermediate-band features. 
Our results substantially expand the known phase space of cubic chiral semiconductors and may guide future efforts in the discovery and design of chiral materials.
\end{abstract}

\maketitle

\section{Introduction}

Chirality is a basic geometric feature defined by the fact that an object cannot be perfectly overlaid with its mirror image. It manifests across a broad range of length scales, from molecular and biological systems to macroscopic structures. In condensed matter physics and materials science, chirality has emerged as an important structural degree of freedom~\cite{bousquet2024structural, FecherReview2023, gomezortiz2024structural, gomezortiz2025pathways}. In crystalline solids, structural chirality requires the absence of all orientation-reversing symmetry operations, including inversion, mirror reflection, and improper rotations. This symmetry condition distinguishes chiral crystals from their achiral counterparts and allows electronic, magnetic, and optical phenomena that are forbidden or more strongly constrained in crystals possessing improper spatial symmetries~\cite{ma2017chiral, ma2019chiral, pop2019main, chang2018topological, wang2025direct}.

Chirality can also combine with band topology to produce unconventional electronic states, including spin-selective transport and topologically protected excitations~\cite{bousquet2024structural, FecherReview2023, FecherReview2022, naaman2012chiral, naaman2015spintronics, yang2025chirality}. Such effects are particularly attractive for low-power spintronic devices, as they enable spin polarization without the need for ferromagnetic layers or external magnetic fields~\cite{sun2024charge, ben2017magnetization, dalum2019theory}. Interesting examples of chirality-driven phenomena include the transition-metal silicides \ce{MnSi} and \ce{Fe_{1-x}Co_xSi}, which crystallize in the chiral B20 structure ($P2_13$, No. 198) and host complex magnetic textures~\cite{muhlbauer2009skyrmion, yu2010realspace, FecherReview2022, altynbaev2022exotic}. Similarly, cubic chiral semimetals such as RhSi and CoSi exhibit unconventional multifold chiral fermions near the Fermi level and extended Fermi-arc surface states~\cite{chang2017unconventional, li2019chiral, hasan2021weyl}.

Among the 22 enantiomorphic space groups, only one pair is cubic: $P4_332$ (No. 212) and its mirror image $P4_132$ (No. 213), which contain fourfold screw axes in addition to 2$_1$ operations. Crystals in these space groups are chiral by symmetry, even when they are built from achiral units. Their point group 432 is non-polar, so pyroelectricity is forbidden. Only a few compounds are experimentally known in these space groups, and most have complex crystal structures. The simplest examples are the binary disilicides \ce{SrSi2} and the high-pressure phase of \ce{BaSi2} (cP12)~\cite{pringle1972structure, evers1974electrical, imai2012structural, bousquet2024structural}. Other binaries, such as \ce{Mg3Ru2} (cP20), \ce{V8C7} (cP60), \ce{K6Sn25} (cP124), \ce{K8Sn25} (cP136), and \ce{RuZn6} (cP252), span a wide range of unit-cell sizes~\cite{FecherReview2022}. Ternary compounds, such as \ce{Mo3Al2C} (cP24), \ce{CsBe2F5} (cP32), \ce{Ag3AuS2} (cP48), \ce{Zn2Ge3O8} (cP52), \ce{Mn3Cu3O8} (cP56), and \ce{LiFe5O8} (cP56), further illustrate the structural complexity typically required to stabilize cubic chirality~\cite{FecherReview2022}. Even so, compounds in these space groups display a broad range of functionalities, including skyrmions above room temperature in $\beta$-Mn-type Co--Zn--Mn alloys~\cite{tokunaga2015new}, high-voltage Li-ion intercalation in the ordered spinel \ce{LiNi_{0.5}Mn_{1.5}O4}~\cite{kim2004comparative}, a spin-liquid state in the hyperkagome antiferromagnet \ce{Na4Ir3O8}~\cite{okamoto2007spin}, and noncentrosymmetric superconductivity in \ce{Li2Pd3B} and \ce{Li2Pt3B}~\cite{yuan2006swave}.

Chiral semiconductors are of particular interest because they couple electronic and optical degrees of freedom: their optical activity gives rise to circular dichroism and to selective interaction with circularly polarized light, which are exploited in photonic and optoelectronic devices~\cite{yan2024structural, bousquet2024structural, schellman1975circular, ahn2023chiral}. Promising examples exist, such as trigonal tellurium, an elemental chiral semiconductor with chirality-dependent spin textures and charge-to-spin conversion~\cite{gatti2020radial, calavalle2022gate}. However, chiral semiconductors remain strongly underrepresented among known inorganic crystals, and their limited chemical and structural diversity is a major bottleneck~\cite{bousquet2024structural}. High-throughput computational studies have recently begun to search for new chiral materials, for example by screening existing databases for chiral topological semimetals~\cite{wang2025exhaustive} and chiral phonon materials~\cite{yang2026symmetry}. Most recently, Song \textit{et al.}~\cite{song2026accelerating} generated about 20 million random ternary structures in the 65 Sohncke space groups, relaxed them with a universal machine-learning interatomic potential, and confirmed by DFT more than 260 dynamically stable nonmagnetic chiral crystals, focusing on their topological, nonlinear optical, and superconducting properties. Their search was, however, restricted to ternary compositions and to unit cells with at most 24 atoms, which excludes the cubic chiral prototypes discussed here, whose unit cells typically contain 32 to 56 atoms.

In this work, we address this problem through a systematic computational search for chiral crystals in the cubic space group $P4_332$ (No.~212), combining symmetry-guided structure generation, machine-learning methods~\cite{merchant2023scaling, alexandria}, and density functional theory (DFT). Structures in its mirror image $P4_132$ are obtained by inversion of the atomic coordinates, $(x,y,z)\rightarrow(-x,-y,-z)$, and have identical energies in the absence of external symmetry-breaking fields, so it is sufficient to explore one member of the pair. In total, we report 24\,227 DFT-relaxed cubic chiral structures containing up to five chemical elements. We analyze their thermodynamic stability in terms of the energy above the convex hull and examine the key structural prototypes. We further investigate their electronic properties, including band gaps and band structures. The full dataset is openly available in the \alex{} database for further exploration of chiral properties.

\section{Results}
\subsection {Dataset generation and validation}
\begin{figure}[ht!]
  \centering
  \includegraphics[width=\columnwidth]{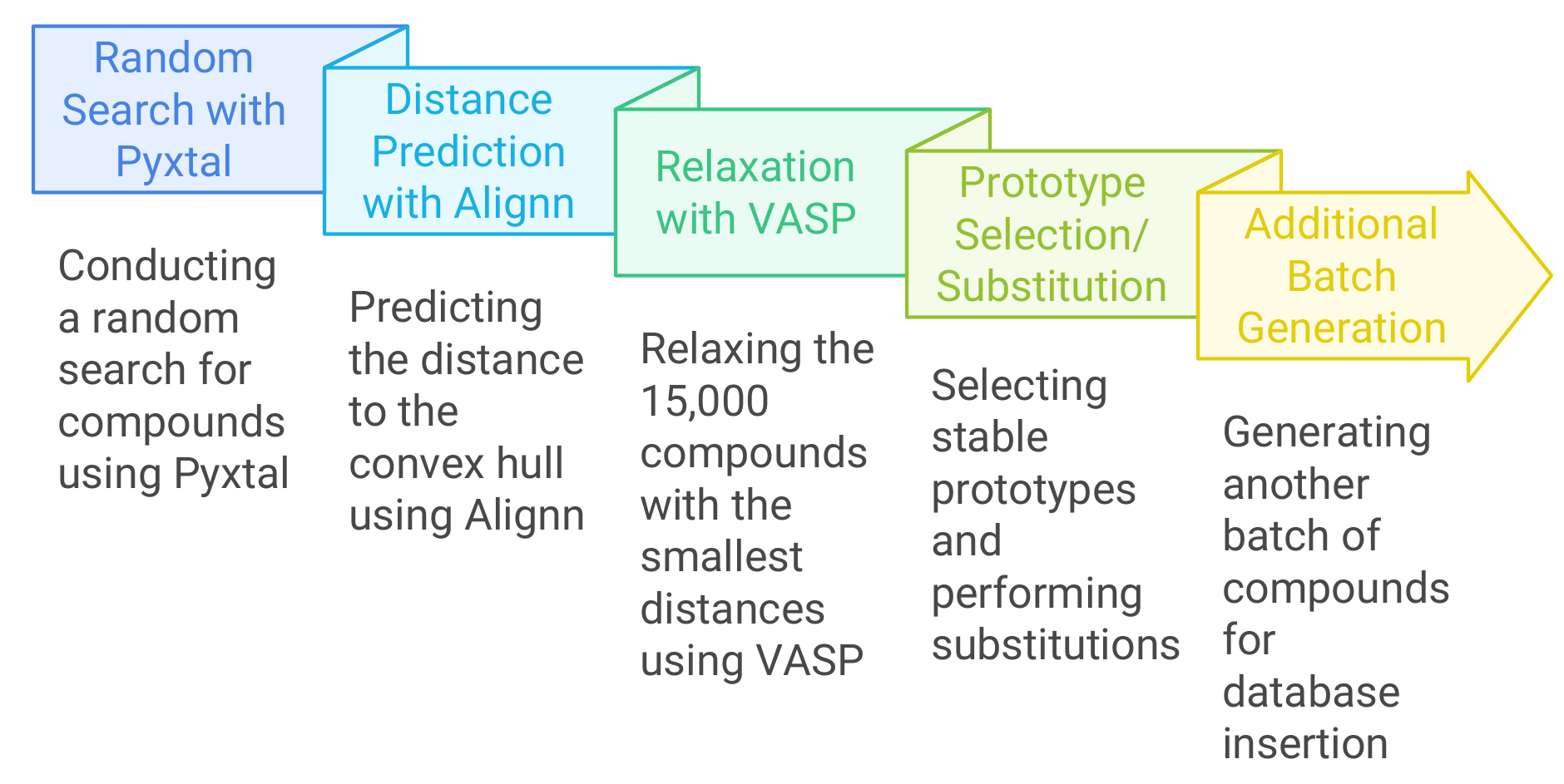}
  \caption{Schematic illustration of the proposed workflow for the structural search of chiral cubic materials. The process includes random structure generation, machine-learning-based geometry optimization, stability estimation via an \textsc{alignn} model, DFT refinement of selected candidates, and iterative element substitution to expand the dataset.}
  \label{fig:workflow}
\end{figure}

\Cref{fig:workflow} illustrates the structure generation workflow, which largely follows our previous approach~\cite{wang2023symmetry, alexandria}. The process began with the generation of 500\,000 random structures by assigning chemical elements to Wyckoff positions while ensuring charge neutrality, using oxidation states from Ref.~\cite{wang2023symmetry}. Although binary compounds were not explicitly excluded, the search primarily favored ternary, quaternary, and quinary compositions. Since these structures were initialized with random Wyckoff positions and lattice parameters, they were far from equilibrium. To refine them, we performed geometry optimization using \textsc{m3gnet}~\cite{m3gnet}, a computationally efficient machine-learning interatomic potential.

To assess stability, we used an \textsc{alignn}~\cite{choudhary2021atomistic} model trained on 3.8 million structures to predict convex hull distances from \textsc{m3gnet}-optimized geometries (the model achieves a mean absolute error of 33~meV/atom on a hold-out test set). Since performing DFT calculations on all structures is impractical, we selected the 15\,000 compounds with the smallest predicted distances to the convex hull for further refinement. These structures were then optimized with DFT as implemented in the VASP package~\cite{vasp1, vasp2} and added to the \alex{} database~\cite{alexandria}. After convergence checks and duplicate removal, 10\,094 unique cubic chiral structures were retained, forming the first batch of relaxed structures.

Using this initial dataset, we identified structural prototypes by anonymous structure matching. A total of 251 prototypes corresponding to (meta-)stable compounds (within 200~meV/atom of the convex hull) were identified. To generate additional candidates, we performed a substitution step as in Ref.~\citenum{wang2021predicting}, choosing similar elements according to the modified Pettifor scale~\cite{glawe2016optimal}. The new candidates then passed through the same workflow of geometry optimization and stability prediction, and the 15\,000 structures with the lowest predicted \ehull{} were selected for DFT calculations. This iterative process yielded an additional 14\,133 compounds after convergence and duplicate checks.

\begin{figure}[ht!]
    \centering
    \includegraphics[width=8cm]{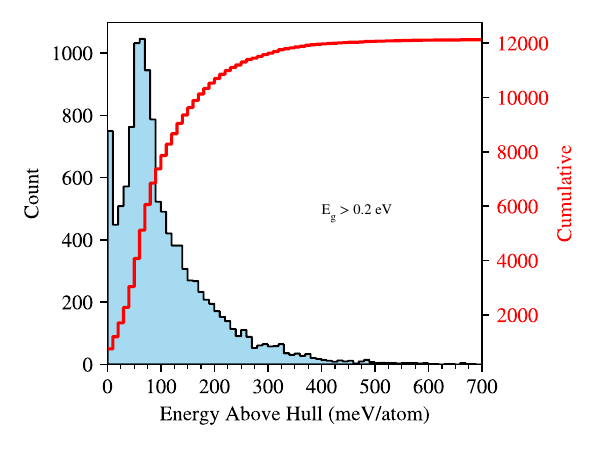}
    \caption{Distribution of the energy above hull for all computed cubic chiral compounds, with a bin width of 10~meV/atom.}
    \label{fig:ehull}

\end{figure}

\begin{figure}[ht!]
    \centering
    \includegraphics[width=8cm]{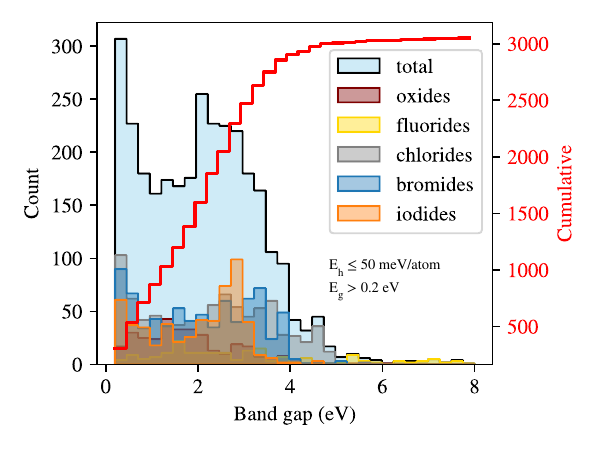}
    \caption{Histogram and cumulative count of the band gaps for all (meta-)stable (\ehull$\le 50$~meV/atom) cubic chiral compounds, using a 25~meV bin width. The 2\,452 entries with a band gap below 0.2~eV are not shown.}
    \label{fig:bandgap}
\end{figure}

At the end of this workflow, 
we obtained a final dataset of 24\,227 unique cubic chiral structures.
\Cref{fig:ehull} presents the histogram of their energy above the convex hull (\ehull). The distribution is not uniform but clearly skewed toward low energies, indicating that the workflow efficiently excludes highly unstable candidates before DFT validation. To focus on semiconductors in the following discussion, 
we selected all systems with a PBE gap $\ge$ 0.2~eV.
We found 368 of them directly on the convex hull, 3\,048 within 50~meV/atom, and 7\,355 within 100~meV/atom above the hull. \Cref{fig:bandgap} shows the band-gap distribution of all semiconducting systems within 50~meV/atom of the hull. Most compounds have a gap below 4~eV, while some systems, mostly fluorides, reach PBE gaps of up to 8~eV. As expected, the gap distribution of the halides shifts toward smaller values from F to I. Interestingly, a portion of the bromides and iodides still have wide PBE gaps of about 3~eV. These are not the usual wide-gap alkali or alkaline-earth bromides and iodides, but mostly compounds of alkali or alkaline-earth metals, transition metals, and Br or I.

\subsection{Structural prototypes and electronic structures}

An advantage of our workflow is its ability to find previously unknown prototypes, as demonstrated before~\cite{wang2023symmetry, alexandria}. To illustrate this, Table~\ref{tab:families} summarizes the families of the most stable compounds (within 50~meV/atom of the convex hull) with more than 50 members. The majority of these prototypes are ternary and quaternary with varying stoichiometries, although several binaries are also common in the dataset. Most of these families occupy the general 24e position, which results in large unit cells with 32 to 56 atoms. This preference for the 24e position is not limited to these families but holds for the entire dataset (see Supplementary Information, SI). Among the most common prototypes in \cref{tab:families} is the \ce{AC2C'2D3X6} family, which contains the experimentally known \ce{Mn3Cu3O8} (or more specifically \ce{CuCu2O2Mn3O6}) and \ce{LiFe5O8} (or \ce{LiFe2O2Fe3O6}). For other experimentally known prototypes, such as \ce{Zn2Ge3O8} (Family \#10), \ce{Mo3Al2C} (Family \#12), and \ce{Mg3Ru2} (Family \#30), we find far fewer (meta-)stable semiconductors (see SI), while for the prototypes of \ce{SrSi2}, \ce{V8C7}, and \ce{CsBe2F5} the computed structures are either unstable or metallic.

In the following, we describe three of these families: the two most common, \ce{ABX6} (\#1) and \ce{AC2D3X6} (\#2), and \ce{AC2C'2D3X6} (\#4), which contains the experimentally known \ce{Mn3Cu3O8} and \ce{LiFe5O8}. For each family we show the structure and the PBE band structure of a representative compound. All families within 50~meV/atom of the hull are listed in the SI.

\begin{table*}[ht!]
\centering
\caption{The nine most common families of cubic chiral semiconductors ($P4_332$) within 50~meV/atom of the convex hull. The elements in 4a, 4b, 8c, 12d, and 24e are denoted as A, B, C, D, and X, respectively. If the same Wyckoff position is occupied more than once in a family formula, the species are distinguished by $'$, $''$, etc. For example, D and D$'$ denote two distinct occupied 12d positions. }
\label{tab:families}
\begin{tabular*}{0.75\textwidth}{@{\extracolsep{\fill}}llclcc}
\toprule
No. & Formula & N$_\text{sites}$ & Wyckoff  & count & Exp. \\
\midrule
\#1 & \ce{ABX6}        & 32 & 4a, 4b, 24e                & 1243 & -- \\
\#2 & \ce{AC2D3X6}     & 48 & 4a/4b, 8c, 12d, 24e        & 618 & -- \\ %
\#3 & \ce{ABD3X6}      & 44 & 4a, 4b, 12d, 24e           & 266 & -- \\ %
\#4 & \ce{AC2C'2D3X6}  & 56 & 4a/4b, 8c, 8c, 12d, 24e    & 253 & \ce{Mn3Cu3O8}, \ce{LiFe5O8} \\ %
\#5 & \ce{AD3X6}       & 40 & 4a, 12d, 24e               & 141 & -- \\ %
\#6 & \ce{ABC2X6}      & 40 & 4a, 4b, 8c, 24e            & 93  & --  \\ %
\#7 & \ce{C2D3D'3}     & 32 & 8c, 12d, 12d               & 77  &  --  \\  %
\#8 & \ce{AC2X6}       & 36 & 4a/b, 8c, 24e              & 76  &  --  \\  %
\#9 & \ce{AC2C'2X6}    & 44 & 4a/b, 8c, 8c, 24e          & 64  &  --  \\  %
\bottomrule
\end{tabular*}
\end{table*}

\subsubsection{\ce{ABX6}}

\newcommand{\minip}[3]{
        \begin{minipage}{#1\columnwidth}
            \centering
            \includegraphics[height=#2]{#3} 
        \end{minipage}}
\begin{figure}
\begin{tabular}{c c}
     \minip{0.27}{3.0cm}{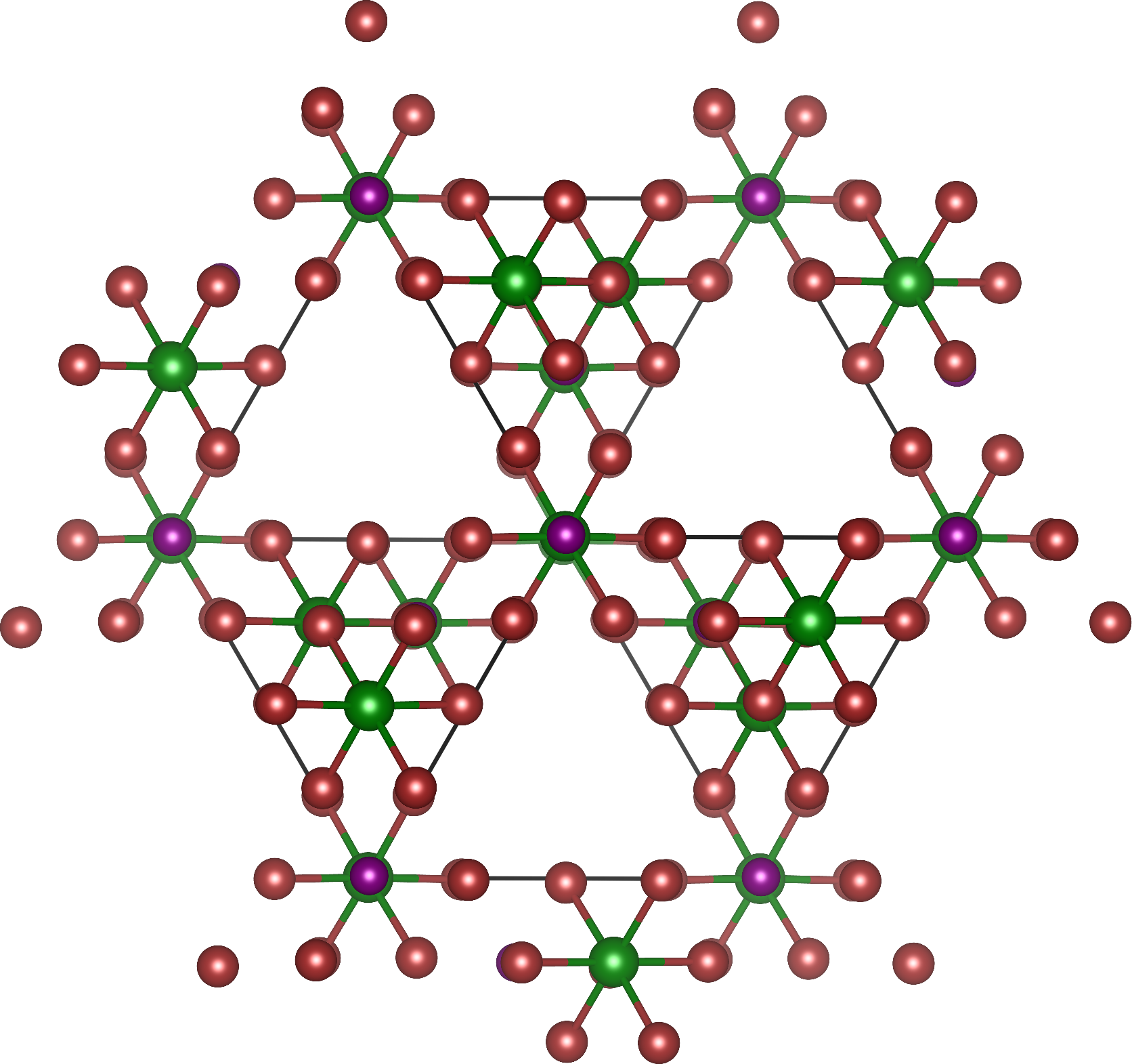} & \hfill
     \minip{0.70}{3.8cm}{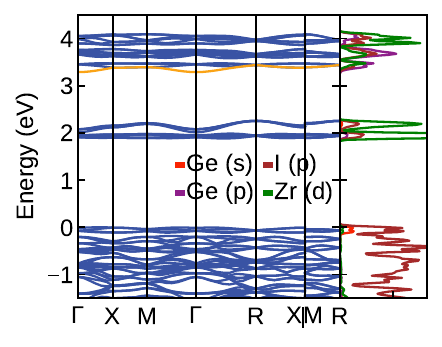} \\
     \hspace{3em}(a) & \hspace{3em}(b) \\ 
\end{tabular}
\caption{(a) The structure of \ce{ZrGeI6} as a representative of the \ce{ABX6} family; Zr atoms are shown in green, Ge in purple, and I in brown. (b) Electronic band structure of \ce{ZrGeI6}; the characteristic dispersive conduction band is highlighted in orange.}
\label{fig:ABX6}
\end{figure}

\Cref{fig:ABX6}(a) shows the \ce{ZrGeI6} structure viewed along the $[111]$ direction. In this motif, the 4a, 4b, and 24e Wyckoff positions are occupied by the A, B, and X elements, respectively, giving the stoichiometry \ce{ABX6}. The structure consists of a network of edge-sharing octahedra, with the A and B atoms at the centers and the X atoms at the vertices. A and B can be the same element, as for Sn in \ce{SnCl3}. 

In this family, the compounds within 50~meV/atom of the hull have either a halogen or hydrogen on the X site. Assuming an oxidation state of $-1$ for the halogens and hydrogen, charge neutrality requires the A and B cations to adopt oxidation states of either \(+1\) and \(+5\), \(+2\) and \(+4\), or both \(+3\). An example of the first case is \ce{Na^{+1}Ta^{+5}Cl6^{-1}}, which has a PBE direct band gap of 3.0~eV. An example of the second case is \ce{Zr^{+4}Ge^{+2}I6^{-1}}, a semiconductor with a PBE band gap of 1.9~eV. An example of the third case is \ce{In^{+3}Ir^{+3}Cl6^{-1}}, which has a PBE direct band gap of 2.3~eV. We note that, in this motif, the general 24e coordinates become highly symmetric after geometry optimization and nearly identical to two sets of 12d positions with different free parameters. We therefore merge the latter case into this family and denote it as \ce{ABD3D'3}. Furthermore, A and B can be identical, for example Bi, resulting in the compound \ce{Bi2I3Cl3}, which lies only 10~meV/atom above the hull and has a direct band gap of 2.53~eV. 

In \Cref{fig:ABX6}(b), we present the band structure of \ce{ZrGeI6}, which is semiconducting with a PBE indirect band gap of 1.88~eV. The conduction bands split into two distinct manifolds separated by a secondary energy gap of 1.03~eV. This feature is characteristic of intermediate-band semiconductors (IBSCs)~\cite{luque2012understanding}, in which intermediate bands (IBs) lie within the band gap. While such IBs are typically engineered through quantum nanostructures, highly mismatched alloys, or high concentrations of impurities, only a limited number of pristine semiconductors show intrinsic IB features~\cite{ghosh2016colloidal, mantella2018synthesis}. Following the IBSC notation, we denote the two sub-gaps in the subsequent band-structure analyses as \El{} (from the valence band to the IB) and \Eh{} (from the IB to the conduction band).
We emphasize, however, that the semiconductors studied here are predominantly charge balanced, so that the true fundamental gap is either \El{} or \Eh{}, depending on the position of the Fermi level.

In \ce{ZrGeI6}, both the valence band maximum (VBM) and the bottom of the IBs are strongly localized, yielding large effective masses for holes (\mh) and electrons (\me) of 25.9 and 39.5 times the electron rest mass (\mer), respectively. As for conventional semiconductors, the projected density of states (PDOS) shows that the valence bands derive mainly from Ge and I $p$ states, while the intermediate and conduction bands have larger contributions from cation orbitals (Zr $d$). The Zr $d$ states dominate the intermediate bands, analogous to deep defect levels induced by $d$-metal doping in semiconductors.

The conduction band minimum (CBM) (highlighted in orange in \cref{fig:ABX6}(b)) shows a pronounced free-electron-like parabolic character, which is uncommon in conventional semiconductors. Despite this parabolic shape, the curvature is modest, resulting in \me{} = 4.5~\mer. As discussed below, we observe this characteristic parabolic band in nearly all representative structures, suggesting a correlation with the chiral nature of the crystal structure.

\subsubsection{\ce{AC2D3X6}} %

\begin{figure}
\begin{tabular}{c c}
     \minip{0.27}{3.0cm}{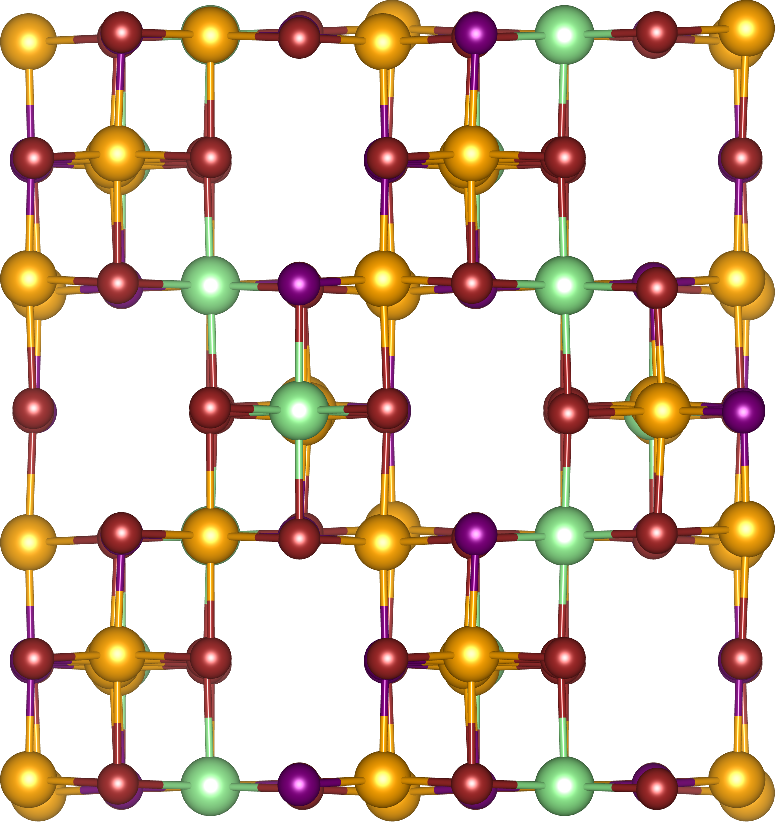} & \hfill
     \minip{0.70}{3.8cm}{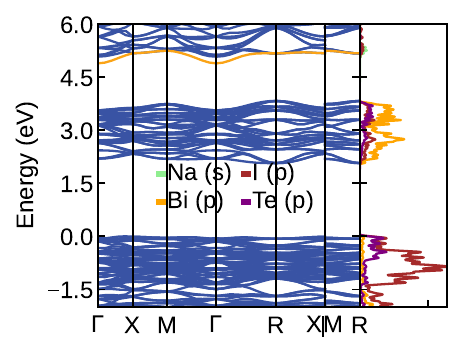} \\
     \hspace{3em}(a) & \hspace{3em}(b) \\ 
\end{tabular}
\caption{(a) The structure of \ce{NaTe2Bi3I6} as a representative of the \ce{AC2D3X6} family; Na, Te, Bi, and I atoms are shown in teal, purple, orange, and brown, respectively. (b) Electronic band structure of \ce{NaTe2Bi3I6}; the characteristic dispersive conduction band is highlighted in orange.}
\label{fig:AC2D3X6}
\end{figure}

\Cref{fig:AC2D3X6}(a) shows the \ce{NaTe2Bi3I6} structure viewed along the $[110]$ direction, illustrating the structural motif of the common \ce{AC2D3X6} family. The structure, with 48 atoms per unit cell, forms a network of edge- and face-sharing octahedra. The A and D cations occupy the centers of the octahedra at the 4a and 12d positions, respectively. The A site is typically occupied by an alkali metal, while the D site is often a rare earth or transition metal in the \(+3\) or \(+4\) oxidation state. The C and X anions are located at the vertices of the octahedra, occupying the 8c and 24e positions. The X atoms are always halogens, while C can be a chalcogen or a halogen. When both the 8c and 24e positions are occupied by halogens, the true ground state is likely disordered, owing to the chemical similarity of the halogens. Additionally, the A and D sites can host the same element, for example Ni$^{2+}$ in \ce{Ni2IBr3}. This family can also have the 4b instead of the 4a position occupied (\ce{BC2D3X6}), which can be seen as the \ce{AC2D3X6} structure with a $(1/2,1/2,1/2)$ translation.

\ce{NaTe2Bi3I6} has an indirect band gap. A similar splitting of the conduction bands is observed, with intermediate bands extending from 2.1 to 3.6~eV, which defines two distinct sub-gaps \El{} and \Eh{} of 2.07 and 1.06~eV, respectively. The VBM consists mainly of hybridized Te and I $p$ states and shows little dispersion, resulting in a large hole effective mass of 12.5~\mer. In contrast, the intermediate bands have significant Bi $p$ contributions and are much more dispersive, with an electron effective mass at the IB minimum of 0.86~\mer. We again observe the characteristic free-electron-like band at the CBM, derived mainly from Na $s$ states, with a small effective mass of \me{} = 0.53~\mer{} that suggests a high electron mobility. These features make this system an interesting candidate for optoelectronic applications, such as intermediate-band photovoltaics and photocatalysis, where efficient carrier transport and multi-band optical transitions are advantageous.

\subsubsection{\ce{AC2C'2D3X6}}%

\begin{figure}
\begin{tabular}{c c}
     \minip{0.27}{3.0cm}{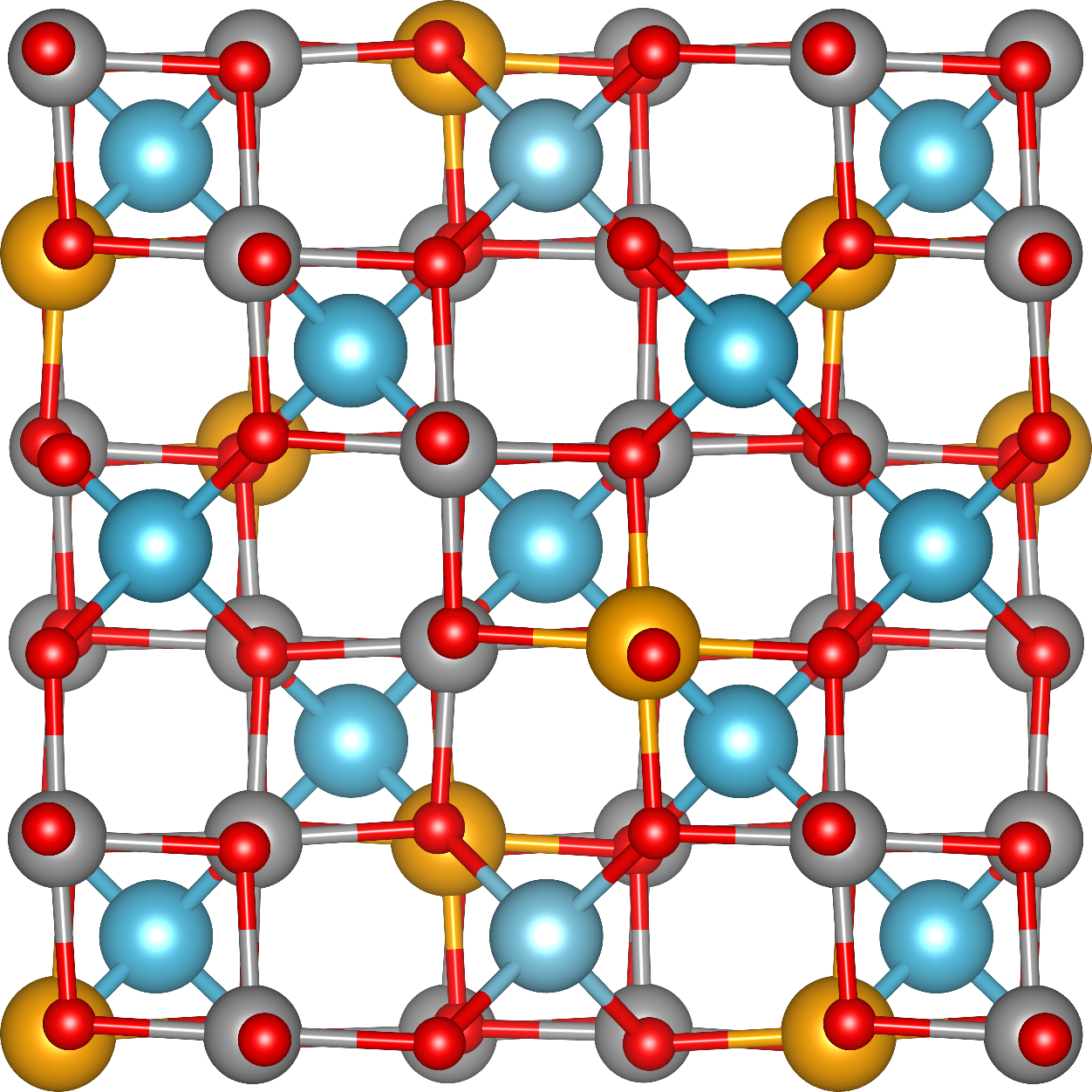} & \hfill
     \minip{0.70}{3.8cm}{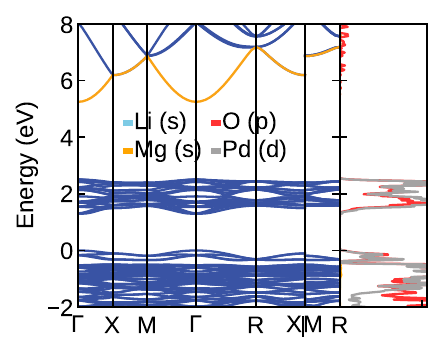} \\
     \hspace{3em}(a) & \hspace{3em}(b) \\ 
\end{tabular}
\caption{(a) The structure of \ce{MgLi2Pd3O8} as a representative of the \ce{AC2C'2D3X6} family; Mg, Li, Pd, and O atoms are shown in orange, light blue, gray, and red, respectively. (b) Electronic band structure of \ce{MgLi2Pd3O8}; the characteristic dispersive conduction band is highlighted in orange.}
\label{fig:AC2C'2D3X6}
\end{figure}

\Cref{fig:AC2C'2D3X6}(a) shows the motif of the common \ce{AC2C'2D3X6} family viewed along the $[100]$ axis, with 56 atoms in the unit cell. This motif is similar to that of the \ce{AC2D3X6} family, forming a network of edge- and face-sharing octahedra, but with additional tetrahedra centered on the extra 8c site. The chemistry also differs slightly: the A site (4a) is typically occupied by a divalent cation, the C site often by an alkali metal, the D site (12d) by a tri- or tetravalent cation, and the X (24e) and C$'$ (8c) sites by a divalent anion. In this study, X and C$'$ are always oxygen: although replacing oxygen with another chalcogen is permitted under the alchemical substitution rules, no compounds of this family with other chalcogens were found within 50~meV/atom of the hull.

An example of this family is \ce{Mg^{+2}Li^{+1}2Pd^{+4}3O^{-2}8}, which has a PBE band gap of 1.3~eV. In some cases, the C cation can also be a divalent transition metal such as Mn, Co, Fe, or Zn. For instance, \ce{CuZn2Co3O8} belongs to this category and has a PBE band gap of 0.5~eV. As for the families discussed above, the composition simplifies in certain cases: the A and C cations can be the same, as in \ce{Co3Rh3O8}; A and D can be the same element, as in \ce{LiMn2O4}; and C and D can be identical, as in \ce{CrFe5O8}. Finally, the 4b instead of 4a position can be occupied, resulting in an alternative \ce{BC2C'2D3X6} motif that can be seen as the \ce{AC2C'2D3X6} motif with a $(1/2,1/2,1/2)$ translation.

The calculated band structure of \ce{MgLi2Pd3O8} (\Cref{fig:AC2C'2D3X6}(b)) also shows intermediate-band character. The VBM and the IBs around 2~eV originate mainly from O $p$ and Pd $d$ states, with moderate effective masses at these band edges (\mh{} and \me{} values of 4.1 and 1.5~\mer, respectively). The CBM, derived mainly from O orbitals, again displays free-electron-like behavior, with a low electron effective mass of 0.5~\mer, which could benefit optoelectronic applications of this system.

An interesting common feature of the representative compounds is the presence of a fairly dispersive, nearly parabolic conduction band which is separated from the lower lying electronic states. This feature is present in a number of different structural families and the orbital character of the dispersion appears to be a function of composition, implying that it is not due to a single atomic species or a single type of orbital. It may well be a more general effect arising from the interplay of local coordination, orbital hybridisation and the connectivity of the underlying crystal structure. The fact that it occurs in such a variety of prototypes is of course very interesting in the context of cubic chiral symmetry, though the present calculations do not establish any direct link between the structural chirality and the nature of the band dispersion. A proper comparison with structurally related achiral compounds would be required to isolate the effect of chirality. It does however point to a potentially useful electronic property of a number of the cubic chiral semiconductor families and does serve to highlight the need to investigate the microscopic origin of the phenomenon.

\subsection{Disorder and chirality}

The structures discussed above are fully ordered and were computed at zero temperature. In real crystals, chemically similar species can mix on different Wyckoff positions, either in equilibrium at the synthesis temperature or frozen in during cooling. This is most likely when two sites hold ions of similar size and charge, such as the halogens on the 8c and 24e positions of the \ce{AC2D3X6} family, or cations of similar radius on the octahedral sites. Disorder destroys chirality only if the averaged structure, with the mixed sites occupied by an average species, has an improper symmetry operation.

To check this for the three families discussed above, we replaced the species that can mix by a single species and determined the space group of the resulting structure with \textsc{spglib}~\cite{spglib}. In \ce{MgLi2Pd3O8} (family \#4), equal occupation of the octahedral 4a and 12d sites gives the centrosymmetric space group $Fd\bar{3}m$ (No.~227) already for a tolerance of 0.1~\AA. This family is an ordered spinel, and its chirality comes entirely from the 1:3 cation order on the octahedral sublattice, so cation disorder restores the achiral spinel structure. This agrees with experiment: \ce{LiNi_{0.5}Mn_{1.5}O4} is known both as an ordered $P4_332$ and as a disordered $Fd\bar{3}m$ phase~\cite{kim2004comparative}, and the degree of Li/Fe order on the octahedral sites of \ce{LiFe5O8} films depends on the growth conditions~\cite{boyraz2011structural}. In contrast, mixing Zr and Ge on the 4a and 4b sites of \ce{ZrGeI6} (family \#1) gives the chiral space group $I4_132$ (No.~214), and \ce{NaTe2Bi3I6} (family \#2) remains in $P4_332$ even when both Na/Bi and Te/I are mixed, for tolerances up to 0.5~\AA. In these two families the occupied positions themselves form a chiral framework, and chirality survives substitutional disorder. Experimental examples of this behavior are the $\beta$-Mn-type alloys, which are chiral despite random site occupations~\cite{FecherReview2022}, and \ce{K8Sn25}, which crystallizes in $P4_132$ with partially occupied K sites in the channels of the Sn framework~\cite{zhao1994zintl}.

Finally, even when the structure is chiral, a crystal can grow with domains of both handedness (racemic twinning), which cancel macroscopic chiral responses such as optical activity. Experiments that probe these responses therefore require single-domain crystals or films.

\section{Conclusion}

In summary, by combining symmetry-constrained structure generation with machine-learning filtering and DFT relaxation, we performed a high-throughput search for cubic chiral compounds in space group $P4_332$ (No. 212). In total, we obtained 24\,227 DFT-relaxed cubic chiral structures and identified 368 semiconducting compounds on the convex hull and 3\,048 within 50~meV/atom.

Across the (meta-)stable subset, the most frequent prototype families are dominated by the occupation of the 24e Wyckoff position, leading to relatively large unit cells (typically 32--56 atoms) and recurrent polyhedral networks, most commonly edge- and face-sharing octahedra, often supplemented by tetrahedral units or bridging atoms. Representative electronic structures show a wide range of band gaps (within the limits of PBE-level screening) and frequently display localized band edges with heavy effective masses, together with a dispersive, free-electron-like conduction-band minimum; in several families we also observe intrinsic intermediate-band-like manifolds that may be relevant for optoelectronic applications. Cation disorder would restore the centrosymmetric spinel structure in the ordered-spinel family, whereas the two most common families remain chiral under substitutional disorder.

The resulting dataset substantially expands the currently known chemical space of cubic chiral semiconductors and provides a set of stable and near-stable prototypes for targeted experimental synthesis and higher-fidelity electronic-structure follow-up (e.g., SOC-inclusive calculations and beyond-PBE gap assessment).

\section{Methodology}

\subsection{Structure generation and machine-learning models}
Random structures were generated with \textsc{pyxtal}~\cite{pyxtal}, and their compositions were screened with the rules implemented in \textsc{smact}~\cite{smact}: only compounds that are charge neutral for the experimentally most common oxidation states~\cite{wang2023symmetry} and that pass the Pauling electronegativity test were retained. Duplicate structures were removed with the structure matcher of \textsc{pymatgen}~\cite{pymatgen}, using its default tolerances.
We used the pretrained \textsc{m3gnet} model distributed with \textsc{m3gnet}~\cite{m3gnet} to pre-optimize the geometries and a retrained \textsc{alignn} model to predict the distance to the hull from the \textsc{m3gnet} relaxed structures.
The training of the \textsc{alignn} model used a set of 4.4 million datapoints. From this dataset we split 300 thousand entries for validation and 300 thousand for testing, while the remaining 3.8 million entries were used for training. We performed 200 epochs, and the best model according to the validation error was selected for the screening. The MAE of this \textsc{alignn} model that predicted the distance to the hull from \textsc{m3gnet}-relaxed structures was 33~meV/atom.

\subsection{DFT calculations}
All geometry optimizations and total-energy calculations were performed with the Vienna Ab initio Simulation Package (\textsc{vasp})~\cite{vasp1, vasp2}. Brillouin zone integration was performed using $\Gamma$-centered $k$-meshes with densities of 1000 k-points per reciprocal atom and with 0.2~eV smearing using first order Methfessel-Paxton method. For magnetic materials, we performed collinear calculations with ferromagnetic states as initial configurations. We used the projector augmented wave (PAW) setups~\cite{paw, paw2} in \textsc{vasp} 5.2, with an energy cutoff of 520~eV. We set the convergence criteria of the forces to be less than 0.005~eV/\AA. Exchange and correlation were described using the Perdew-Burke-Ernzerhof (PBE) functional~\cite{pbe}, supplemented with Hubbard $U$ corrections for transition metal oxides and fluorides containing Co, Cr, Fe, Mn, Mo, Ni, V, or W, with $U$ values for $d$-states being 3.32, 3.7, 5.3, 3.9, 4.38, 6.2, 3.25, and 6.2 eV, respectively. The electronic self-consistency was converged to $10^{-7}$~eV. Spin--orbit coupling was not included. Calculations that did not converge were discarded, and distances to the convex hull were evaluated with \textsc{pymatgen}~\cite{pymatgen} using the convex hull of the \textsc{alexandria} database. All computational parameters, convergence criteria, and pseudopotential selections were configured to maintain consistency with the \textsc{alexandria} database~\cite{HCGAT, wang2023symmetry, alexandria, cavignac2026ai}.

The study's high throughput means we are walking a fine line between computational accuracy and the sheer scale of chemical and structural space we're trying to search. The ML predictions are used purely to prioritize candidate structures for DFT (first-principles) calculations. All energies, electronic structures, and distances to the convex hull reported are from DFT. The initial screen could remove some potentially stable compounds due to uncertainties in the ML predictions, and the resulting set should be treated as a targeted search rather than a complete enumeration of cubic chiral semiconductors. The calculated convex hull distances measure thermodynamic stability only with respect to the competing phases in the reference database and within the approximations of the exchange-correlation functional and magnetic configurations we used.  The lack of spin-orbit coupling and the use of collinear magnetic structures may have affected the electronic and energetic properties of some compounds (particularly those with heavy elements or magnetic transition metals). Dynamical stability and the potential for lower symmetry distortions have not been fully explored either. That said, these caveats do not diminish the candidates identified; the point is that the selected compounds will need to be validated experimentally before we can be confident in their actual stability and functionality.

\section*{Data availability}
The data that support the findings of this study, including all relaxed crystal structures, total energies, distances to the convex hull, and band gaps, are openly available in the \alex{} database at \url{https://alexandria.icams.rub.de/}.

\begin{acknowledgments}
M.A.L.M. was supported by a collaboration between the Kavli Foundation, Klaus Tschira Stiftung, and Kevin Wells, as part of the SuperC collaboration, and by the Simons Foundation through the Collaboration on New Frontiers in Superconductivity (Grant No. SFI-MPS-NFS-00006741-10). A.H.R. acknowledges support from the West Virginia Higher Education Policy Commission through the Research Challenge Grant Program (Award No. RCG 23-007, 2022). The authors gratefully acknowledge the computing time made available to them on the high-performance computers Noctua and Otus at the NHR Center Paderborn Center for Parallel Computing (PC2). Work at West Virginia University was supported by the grant NSF-OAC-2513657 funded by the National Science Foundation. This work utilized Bridges2 and Expanse at the Pittsburgh Supercomputer (Bridges2) and the San Diego Supercomputer Center (Expanse) through allocation DMR140031 from the Advanced Cyberinfrastructure Coordination Ecosystem: Services \& Support (ACCESS) program, which is supported by the National Science Foundation through grants 2138259, 2138286, 2138307, 2137603, and 2138296. 
\end{acknowledgments}

\bibliography{bib}

\clearpage
\onecolumngrid
\newcount\supppage
\AddToShipoutPictureBG{%
  \supppage=\value{page}\advance\supppage by -9
  \ifnum\supppage>0 \ifnum\supppage<26
    \AtPageLowerLeft{\includegraphics[page=\the\supppage]{supplementary.pdf}}%
  \fi\fi}
\pagestyle{empty}
\newcount\suppcount
\suppcount=0
\loop
  \null\clearpage
  \advance\suppcount by 1
\ifnum\suppcount<25 \repeat

\end{document}